\documentclass[a4paper]{jpconf}
\usepackage{graphicx}
\usepackage{bm}
\usepackage{braket}
\usepackage{amssymb}
\usepackage{amsmath}
\usepackage{color}
\usepackage{txfonts}
\usepackage{bm}
\usepackage{ulem}

\begin{document}
\title{Surface and interface effects on a magnetic Chern insulator}

\author{Ryo Ozawa$^1$, Masafumi Udagawa$^1$, Yutaka Akagi$^2$, and Yukitoshi Motome$^1$}

\address{$^1$Department of Applied Physics, University of Tokyo, Bunkyo, Tokyo 113-8656, Japan}
\address{$^2$Okinawa Institute of Science and Technology, Onna, Okinawa 904-0412, Japan}

\ead{ozawa@aion.t.u-tokyo.ac.jp}

\begin{abstract}
Effects of an open surface on a magnetic Chern insulator are investigated in comparison with those of an interface to a capping magnetic layer.
In magnets, an open surface often perturbs the magnetic order by a reconstruction of the magnetic moment directions near the surface. 
On the other hand, in topological insulators, it leads to the formation of topologically protected surface states.
These two contrasting effects may coexist in magnetic Chern insulators, which give rise to nontrivial surface reconstruction.
For instance, the chiral edge current is largely enhanced by the edge reconstruction in a two-dimensional magnetic Chern insulator realized in a quarter-filled Kondo lattice model on a triangular lattice.
We here show that the edge reconstruction can be described semiquantitatively by a simple junction model between the bulk topological magnetic state and a ferromagnetic capping layer. 
We further clarify how the chiral edge current is affected by the magnetic structure in the capping layer. 
Our results indicate that the topological edge state can be controlled magnetically through the junctions.
\end{abstract}

\section{Introduction}\label{sec:intro}

A surface is one of the central subjects in condensed matter physics. 
An open surface changes the spatial symmetry of the system and perturbs the bulk electronic states, which sometimes brings about novel surface states qualitatively different from the bulk one~\cite{PhysRev.115.869,PhysRevLett.43.43,PhysRevLett.81.1953}. 
In particular, in topological insulators, the surface provides a stage where the nontrivial topological  
nature manifests itself, in the form of topologically protected surface states~\cite{PhysRevB.25.2185,PhysRevB.48.11851,RevModPhys.82.3045}. 
Such topological surface states exhibit a number of interesting properties, such as a dissipationless chiral edge current, which are of special interest for device applications as well as the fundamental science. 

\begin{figure}[h]
	\begin{center}
	\includegraphics[width=15cm]{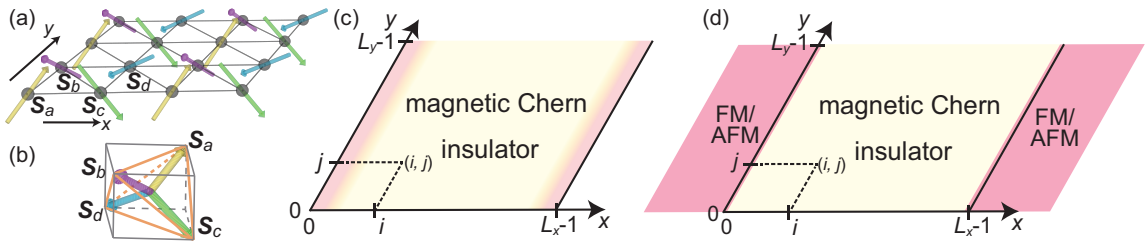}
	\caption{\label{fig:schematics} 
	Schematics of (a) a four-sublattice spin order on a triangular lattice, 
	(b) spin directions of the four-sublattice order which form a tetrahedron, 
	(c) a MCI with open edges in the $x$ direction,  
	and (d) a MCI with capping magnetic layers. 
	In (c), spin configurations are optimized to minimize the free energy~\cite{JPSJ.83.073706}.
	See the text for details. 
	}
	\end{center}
\end{figure}

Among many systems with a topologically nontrivial character, magnetic Chern insulators (MCIs) have recently attracted special interest. 
The MCIs are magnetically ordered insulators that possess the electronic structure characterized  
by a nonzero Chern number~\cite{PhysRevB.62.R6065,PhysRevLett.87.116801}. 
The topologically nontrivial nature usually originates in the so-called spin Berry phase carried by a noncoplanar magnetic long-range order. 
Such noncoplanar magnetic structures are often characterized by the spin scalar chirality: $\chi_{lmn} \propto (\bm{S}_l\times\bm{S}_m)\cdot\bm{S}_n$ defined for three spins $\bm{S}_l$, $\bm{S}_m$, and $\bm{S}_n$. 
A typical example of MCIs was recently reported in the Kondo lattice model on a triangular lattice near 1/4 and 3/4 fillings~\cite{PhysRevLett.101.156402, JPSJ.79.083711,PhysRevLett.105.266405}. 
This MCI is accompanied by a four-sublattice noncoplanar magnetic order with a nonzero scalar chirality, as shown in Figs.~\ref{fig:schematics}(a) and \ref{fig:schematics}(b). 
The instability toward such peculiar magnetic ordering was discussed from the viewpoint of the Fermi surface properties~\cite{PhysRevLett.101.156402, PhysRevLett.108.096401, PhysRevB.90.060402}.
Different types of MCIs were also reported in other lattice systems~\cite{PhysRevLett.109.166405,barros2014novel}. 

MCIs exhibit the topological quantum Hall effect, associated with the nonzero Chern numbers. 
In the quantum Hall insulating state, the surfaces become metallic and retain the chiral edge currents. 
However, in contrast to ordinary (nonmagnetic) Chern insulators, the emergence of such topologically protected edge states is not trivial. 
This is because, in general, the magnetic structure is perturbed by the spatial symmetry breaking by the surface. 
Once this occurs in MCIs, the topological nature is also perturbed by the surface, which might significantly affect the edge states.

In fact, in the previous study~\cite{JPSJ.83.073706}, the authors found that, for the chiral edge states in the MCI on the triangular lattice, the four-sublattice noncoplanar magnetic order is substantially reconstructed and ferromagnetic (FM) correlations develop near the edges. 
Surprisingly, the FM correlations  enhance the total amount of chiral edge current up to almost twice, rather than suppress it. 
It is interesting how the edge magnetic reconstruction leads to the enhancement of the chiral edge current.
However, the precise description of the reconstructed edge state is complicated: it is obtained only after the minimization of many-body free energy by optimizing spin configurations near the edges. 
It is desirable to establish an effective model, which captures the essential physics of the edge reconstruction in the MCI. 

In this paper, we show that the essential character of reconstructed edge states can be described by a simple junction model between the bulk four-sublattice ordered state and a FM layer.
The electronic state near the interfaces in the junction system gives a good approximation to
the edge states in the optimized spin configuration: 
the total chiral current is largely enhanced by the FM junction. 
Furthermore, we find that the total chiral current is strongly affected by the magnetic structure of the capping layers; for instance, it is suppressed substantially by an antiferromagnetic (AFM) junction. Through the careful analysis of electronic band structures of the edge states, we discuss the origin of the enhanced chiral edge current in terms of a variant of the double-exchange mechanism.

\section{Model and method}\label{sec:model}
In this study, we focus on the MCI with the four-sublattice noncoplanar spin configuration on a triangular lattice as schematically shown in Figs.~\ref{fig:schematics}(a) and \ref{fig:schematics}(b). 
This is realized in the Kondo lattice model at 1/4 filling~\cite{JPSJ.79.083711,PhysRevLett.105.266405}, whose Hamiltonian is given by
\begin{equation}
\hat{\mathcal{H}} = -t \sum_{\braket{l, m}}\sum_s(\hat{c}^\dagger_{ls}\hat{c}_{ms} + {\rm h.c.}) - J_{\rm H}\sum_l\hat{\bm{s}}_l\cdot{\bm{S}}_l.
\label{Hamiltonian}
\end{equation}
Here, the first term represents hopping of itinerant electrons, where $\hat{c}^\dagger_{ls}(\hat{c}_{ls})$ denotes the creation (annihilation) operator of an itinerant electron on site $l$ with spin $s=\uparrow, \downarrow$, $t$ is the transfer integral, and the sum $\braket{l,m}$ is taken over nearest neighbor sites on a triangular lattice. 
The second term is the on-site exchange interaction between localized spins $\bm{S}_l$ and itinerant electron spins $\bm{\hat{s}}_l = \sum_{s, s'}\hat{c}^\dagger_{ls}\bm{\sigma}_{ss'}\hat{c}_{ls'}$ ($\bm{\sigma}$ is the vector representation of the Pauli matrix); 
$J_{\rm H}$ denotes the coupling constant. 
Hereafter, we assume $\bm{S}_l$ to be a classical vector with $|\bm{S}_l|$ = 1, and take $t=1$ as an energy unit and the lattice constant as a length unit.  
Also, we denote the coordinate of site $l$ by ($i$, $j$) measured from the origin defined on the left edge or interface 
[see Figs.~\ref{fig:schematics}(c) and \ref{fig:schematics}(d)]. 
In the following calculations, the system size is taken to be $L_x \times L_y$ sites with an open (periodic) boundary condition in the $x$ ($y$) direction.
All the following calculations are done at 1/4 filling and $J_{\rm H}=3$.

The four-sublattice noncoplanar order in the MCI shown in Figs.~\ref{fig:schematics}(a) and \ref{fig:schematics}(b) is given by the spin configurations,
$\bm{S}_{(2m,     2n    )} = \frac{1}{\sqrt{3}}(  1,  1,  1)$, 
$\bm{S}_{(2m,     2n+1)} = \frac{1}{\sqrt{3}}( -1, -1,  1)$,  
$\bm{S}_{(2m+1, 2n    )} = \frac{1}{\sqrt{3}}(  1, -1, -1)$, and  
$\bm{S}_{(2m+1, 2n+1)} = \frac{1}{\sqrt{3}}( -1,  1, -1)$; 
$m$ and $n$ are integers in $0\leq m \leq L_x/2-1$ and $0\leq n \leq L_y/2-1$. 
We denote this perfectly ordered configuration as $\{\bm{S}_l^{\rm 4sub}\}$. 
Meanwhile, the optimized spin configuration in the system with open edges in the $x$ direction [see Fig.~\ref{fig:schematics}(c)] is denoted by $\{\bm{S}_l^{\rm opt}\}$. 
The optimized state was obtained in the previous study by using the Langevin simulation with
the kernel polynomial expansion method at zero temperature;  
for the details, the readers are referred to Ref.~\cite{JPSJ.83.073706}.

In addition to them, we consider the ``junction-type" spin configurations, $\{\bm{S}^{\theta}_l\}$. 
This is defined by replacing each edge ($i=0$ and $i=L_x-1$) of $\{\bm{S}^{\rm 4sub}_l\}$ by a magnetically ordered layer. 
We call the replaced edges the capping layers.
In the capping layers, we assume a twist of spins from $\{\bm{S}^{\rm 4sub}_l\}$ by an angle $\theta$.
Namely, the configuration $\{\bm{S}^{\theta}_l\}$ is given by 
\begin{equation}
\label{eq:S^theta}
	\bm{S}^{\theta}_{(i,j)} = 
	\begin{cases}
	\bm{S}_z\cos(\theta/2) + (-1)^j\bm{S}_{xy}\sin(\theta/2) & (i=0) \\
        -\bm{S}_z\cos(\theta/2) + (-1)^j\bar{\bm{S}}_{xy}\sin(\theta/2) & (i=L_x-1) \\
	\bm{S}^{\rm 4sub}_{(i,j)} & ({\rm otherwise}),
	\end{cases}
\end{equation}
where $\bm{S}_z=(0, 0, 1)$, $\bm{S}_{xy}=\frac{1}{\sqrt{2}}(1, 1, 0)$, and $\bar{\bm{S}}_{xy}=\frac{1}{\sqrt{2}}(1 ,-1, 0)$. 
Note that the spins in the capping layers are aligned ferromagnetically (antiferromagnetically) for $\theta=0(\pi)$, which are denoted by $\{\bm{S}^{\rm FM}_l\} = \{\bm{S}^{\theta=0}_l\}$ and $\{\bm{S}^{\rm AFM}_l\} = \{\bm{S}^{\theta=\pi}_l\}$, respectively. 
We also note that $\{\bm{S}^{\theta}_l\}$ with $\theta=\cos^{-1}(-1/3)\sim0.6\pi$ coincides with $\{\bm{S}_l^{\rm 4sub}\}$. 
In other words, the configuration $\{\bm{S}^{\theta}_l\}$ continuously interpolates the perfect four-sublattice order, FM, and AFM interfaces by changing $\theta$.

For each spin configuration, we diagonalize the fermionic part of the Hamiltonian in Eq.~(\ref{Hamiltonian}), and obtain the electronic state at zero temperature. 
We calculate the electronic band structure and the local electric current density parallel to the edge, $j_{\parallel}(i)$, which is defined as
\begin{equation}
\label{eq:j}
j_\parallel(i) = \frac{1}{2{\rm i}L_y}\sum_{j=0}^{L_y-1}\sum_{s=\uparrow, \downarrow} 
\langle\hat{c}^\dagger_{(i,j)s}\hat{c}_{(i,j+1)s} - {\rm h.c.}\rangle, 
\end{equation}
and compare the results for $\{\bm{S}_l^\theta\}$ and $\{\bm{S}_l^{\rm opt}\}$. 

\section{Results and discussions}\label{sec:results}
\begin{figure}[h]
	\begin{center}
	\includegraphics[width=14cm]{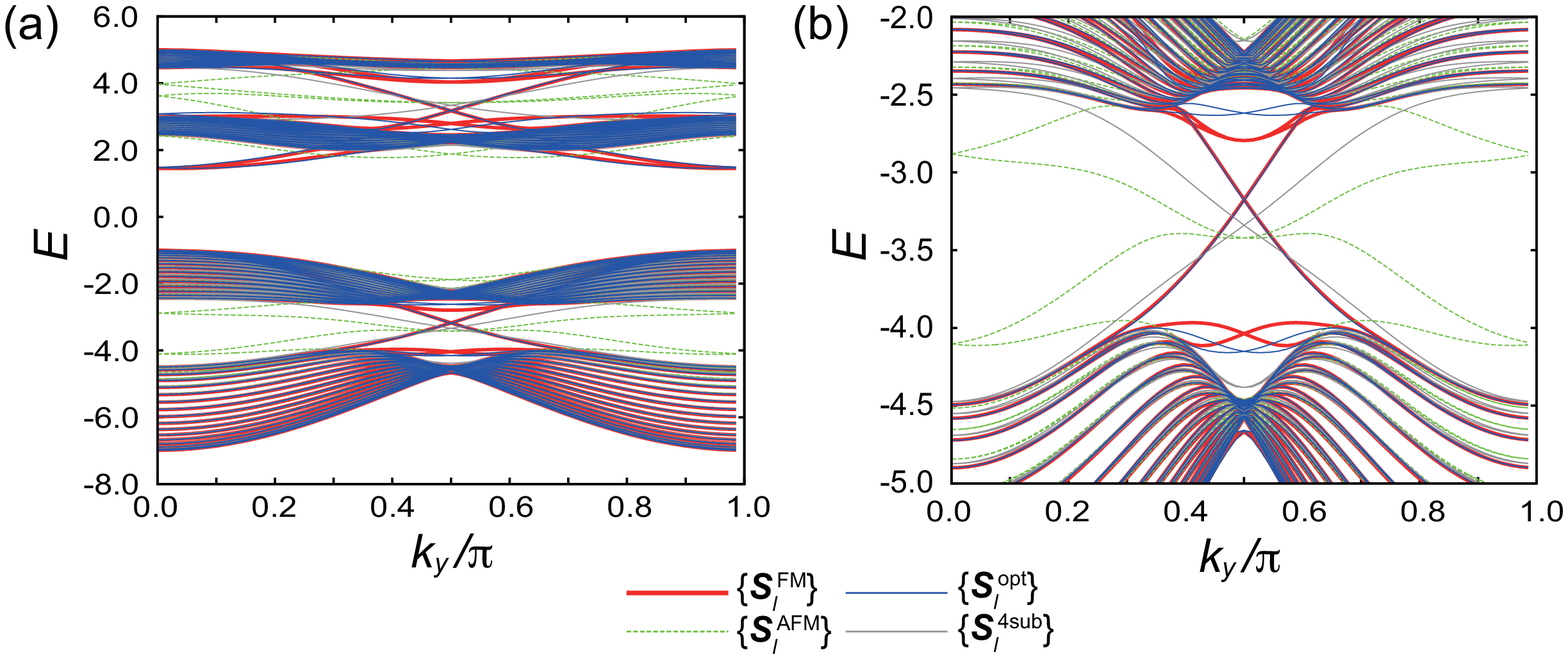}
	\caption{\label{fig:band} 
	Band structures of the MCIs with
	FM capping layers, $\{\bm{S}_l^{\rm FM}\}$, 
	AFM capping layers, $\{\bm{S}_l^{\rm AFM}\}$, 
	the optimized spin configuration for the system with open edges, $\{\bm{S}_l^{\rm opt}\}$, 
	and the four-sublattice order,  $\{\bm{S}_l^{\rm 4sub}\}$. 
	The data are calculated at $J_{\rm H}=3$ and for $L_x\times L_y=34\times136$. 
	(a) shows the overall energy spectra, and
	(b) is an enlarged figure of (a) near the chemical potential for 1/4 filling. 
	The dispersions of the edge states in the bulk gap ($-4.0 \lesssim E \lesssim -2.8$) almost coincide with each other for $\{\bm{S}_l^{\rm FM}\}$ and $\{\bm{S}_l^{\rm opt}\}$. 
	}
	\end{center}
\end{figure}

Figure~\ref{fig:band} shows the electronic band structures as functions of the momentum in the $y$ direction, $k_y$, for the four types of spin configurations, $\{\bm{S}_l^{\rm FM}\}$, $\{\bm{S}_l^{\rm AFM}\}$, $\{\bm{S}_l^{\rm 4sub}\}$, and $\{\bm{S}_l^{\rm opt}\}$:  
Fig.~\ref{fig:band}(a) shows the overall band structure, and Fig.~\ref{fig:band}(b) is the enlarged figure near the chemical potential for 1/4 filling.
Here, we calculate the band structures for the systems with $L_x\times L_y = 34\times 136$ sites; for $\{\bm{S}_l^{\rm opt}\}$, the spin configuration obtained for $34\times 34$ sites is repeated in the $y$ direction.
As shown in Fig.~\ref{fig:band}(a), the band structure is split into four bunches.  
For the perfect four-sublattice order $\{\bm{S}_l^{\rm 4sub}\}$, each bunch forms an isolated band separated by finite energy gaps, if the periodic boundary condition is assumed also for the $x$ direction.
In the presence of open edges, however, the chiral edge states appear so as to traverse the energy gaps corresponding to 1/4 and 3/4 fillings with the crossing points at $k_y=\pi/2$, as shown in Fig.~\ref{fig:band}(a). 
The existence of chiral edge states results from the nontrivial topological property of the four-sublattice ordered state. 
As indicated in the data for $\{\bm{S}_l^{\rm opt}\}$ in Fig.~\ref{fig:band}, albeit the edge reconstruction considerably modifies the dispersions of the chiral edge states, it does not destroy the topologically protected edge states~\cite{JPSJ.83.073706}. 
The situation is similar to the cases of the capping layers for both $\{\bm{S}_l^{\rm FM}\}$ and $\{\bm{S}_l^{\rm AFM}\}$, but the form of the dispersions are distinct between the two cases: 
the dispersions for $\{\bm{S}_l^{\rm FM}\}$ almost coincide with those for $\{\bm{S}_l^{\rm opt}\}$, while the results for $\{\bm{S}_l^{\rm AFM}\}$ look very different from all other cases as shown in Fig.~\ref{fig:band}(b). 
The results suggest that reconstructed edge states in the system with open edges are well described by that with one layer of FM ``skin". 

\begin{figure}[t]
	\begin{minipage}{14pc}
	\includegraphics[width=7cm]{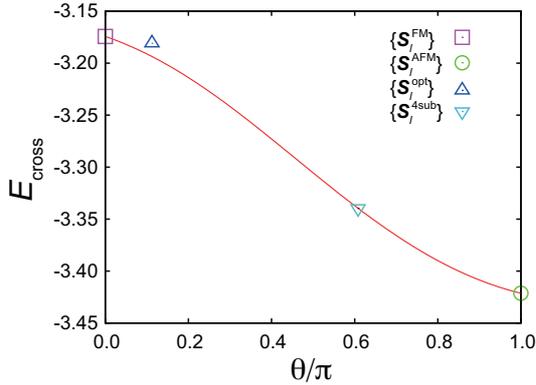}
	\end{minipage}\hspace{2cm}
	\begin{minipage}{8cm}
	\vspace{-1cm}
	\caption{\label{fig:dirac} Energy of the band crossing point near 1/4 filling, $E_{\rm cross}$, 
	as a function of the angle $\theta$ in Eq.~(\ref{eq:S^theta}). 
	The data are calculated at $J_{\rm H} = 3$ and for $L_x\times L_y=34\times136$. 
	The square, circle, upward triangle, and downward triangle indicate $E_{\rm cross}$ for
	$\{\bm{S}_l^{\rm FM}\}$,  $\{\bm{S}_l^{\rm AFM}\}$,
	$\{\bm{S}_l^{\rm opt}\}$, and $\{\bm{S}_l^{\rm 4sub}\}$, respectively.
	See the text for details. 
	}
	\end{minipage}\hspace{20cm}
\end{figure}

In order to further examine how the edge spin configuration affects the chiral edge mode, we consider the crossing energy of chiral modes  at $k_y=\pi/2$ near 1/4 filling. 
Figure~\ref{fig:dirac} shows the crossing energy $E_{\rm cross}(\{\bm{S}_l^\theta\})$ as a function of $\theta$ (red curve).
Note that the data for $\{\bm{S}^{\rm FM}_l\}$, $\{\bm{S}^{\rm AFM}_l\}$, and $\{\bm{S}^{\rm 4sub}_l\}$ are on the red curve at $\theta=0$, $\pi$, and $\sim0.6\pi$, respectively. 
The result indicates that $E_{\rm cross}(\{\bm{S}_l^\theta\})$ monotonically decreases as the edge spin configuration changes from FM to AFM by increasing $\theta$. 
This behavior will be discussed later, in comparison with the enhancement of the chiral edge current.

In Fig.~\ref{fig:dirac}, we also plot the crossing energy for $\{\bm{S}^{\rm opt}_l\}$, $E_{\rm cross}(\{\bm{S}^{\rm opt}_l\})$. 
Here, we define the twist angle $\theta$ for $\{\bm{S}^{\rm opt}_l\}$ by the relative angle between the neighboring spins in the edge layer: $\theta=\cos^{-1}\left(\bm{S}_{(0, j)}^{\rm opt}\cdot\bm{S}_{(0, j+1)}^{\rm opt}\right)$. We note that $\theta$ is insensitive to $j$.
With this definition, $E_{\rm cross}(\{\bm{S}^{\rm opt}_l\})$ is almost on the curve of $E(\{\bm{S}^{\theta}_l\})$, as shown in Fig.~\ref{fig:dirac}. 
This is rather surprising because the reconstruction of the spin configuration from $\{\bm{S}^{\rm 4sub}_l\}$ is not limited to the edge layer~\cite{JPSJ.83.073706}. 
Furthermore, the value of $E_{\rm cross}(\{\bm{S}^{\rm opt}_l\})$ is very close to $E_{\rm cross}(\{\bm{S}^{\rm FM}_l\})$. 
The results indicate quantitatively that the reconstructed edge states for $\{\bm{S}^{\rm opt}_l\}$ are close to those for the simple junction model with $\{\bm{S}^{\rm FM}_l\}$.

\begin{figure}[h]
	\begin{center}
	\includegraphics[width=14cm]{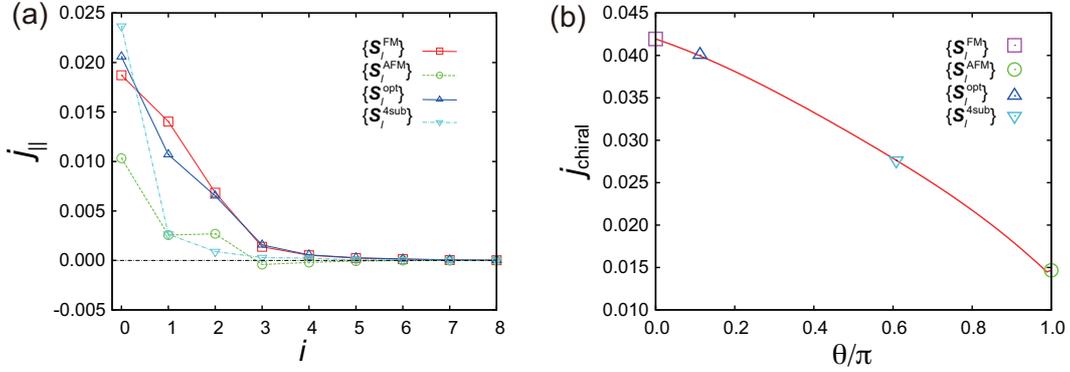}
	\caption{\label{fig:current} 
	(a) Electric current densities $j_{\parallel}$ as functions of the distance from the edge, $i$, under several spin configurations calculated at $J_{\rm H} = 3$ and for $L_x\times L_y=34\times 34$. 
	The squares, circles, upward triangles, and downward triangles indicate $j_{\parallel}$ for 
	$\{\bm{S}_l^{\rm FM}\}$, $\{\bm{S}_l^{\rm AFM}\}$,
	$\{\bm{S}_l^{\rm opt}\}$, and $\{\bm{S}_l^{\rm 4sub}\}$, respectively.
	The lines are the guides for the eye.
	(b) Integrated chiral current $j_{\rm chiral}$ as a function of $\theta$.
	The square, circle, upward triangle, and downward triangle indicate $j_{\rm chiral}$
	for $\{\bm{S}_l^{\rm FM}\}$, $\{\bm{S}_l^{\rm AFM}\}$, 
	$\{\bm{S}_l^{\rm opt}\}$, and $\{\bm{S}_l^{\rm 4sub}\}$, respectively. 
	}
	\end{center}
\end{figure}

In Fig.~\ref{fig:current}(a), we plot spatial modulations of the local current density $j_{\parallel}(i)$ in Eq.~(\ref{eq:j}) for the spin configurations 
$\{\bm{S}_l^{\rm FM}\}$,  $\{\bm{S}_l^{\rm AFM}\}$,  $\{\bm{S}_l^{\rm opt}\}$, and  $\{\bm{S}_l^{\rm 4sub}\}$.
The calculations are done for the systems with $L_x\times L_y = 34\times 34$ sites.
The result shows that the current densities for $\{\bm{S}_l^{\rm FM}\}$ and $\{\bm{S}_l^{\rm opt}\}$ are similar to each other: 
$j_\parallel(0)$ is lower than that for $\{\bm{S}_l^{\rm 4sub}\}$, whereas $j_\parallel(1)$ and $j_\parallel(2)$ are much larger than those for $\{\bm{S}_l^{\rm 4sub}\}$.
The decrease of $j_\parallel(0)$ for $\{\bm{S}_l^{\rm FM}\}$ and $\{\bm{S}_l^{\rm opt}\}$ reflects
the suppression of local spin scalar chirality at the edges because of FM spin correlations (see also Ref.~\cite{JPSJ.83.073706}).

We also plot the integrated chiral current $j_{\rm chiral}$ in Fig.~\ref{fig:current}(b),
 which is defined as $j_{\rm chiral} = \sum_{i=0}^{L_x/2-1} j_\parallel(i)$, as a function of $\theta$ (red curve). 
 As shown in the figure,  $j_{\rm chiral}$ shows almost 1.5 times larger values for  $\{\bm{S}_l^{\rm FM}\}$ and $\{\bm{S}_l^{\rm opt}\}$, 
 compared with that for $\{\bm{S}_l^{\rm 4sub}\}$. 
In contrast, for $\{\bm{S}_l^{\rm AFM}\}$, $j_{\rm chiral}$ is reduced to almost half of the value for $\{\bm{S}_l^{\rm 4sub}\}$.
Comparing the Fig.~\ref{fig:current}(b) with Fig.~\ref{fig:dirac}, we find a similar tendency 
between the $\theta$ dependences of $j_{\rm chiral}$ and $E_{\rm cross}$.
This is understood as follows. 
The amount of the chiral current is roughly proportional to the ``bandwidth" 
of filled edge states, 
i.e., the difference between the chemical potential and the band bottom energy of the edge state. 
This means that $E_{\rm cross}$ gives a good measure for the amount of chiral current $j_{\rm chiral}$, as the band bottom is almost unchanged for different spin configurations. 

The above consideration leads us to associate the increase of $E_{\rm cross}$, or the widening of chiral ``bandwidth", with the double-exchange mechanism~\cite{PhysRev.82.403}. 
In the double-exchange mechanism, the underlying spins are ferromagnetically aligned so as to gain the kinetic energy of itinerant electrons. The gain of the effective kinetic energy in the edge mode for $\{\bm{S}_l^{\rm FM}\}$ and $\{\bm{S}_l^{\rm opt}\}$ implies that a similar mechanism works in both cases; 
the FM correlation suppresses the current density in the outmost layer, but it increases the total amount of edge current by optimizing the kinetic energy in the vicinity of system edges. 
This double-exchange type mechanism may be generic in a wide class of MCIs, and perhaps gives a guiding principle to the general problem of edge reconstruction. 

Sensitivity of chiral edge current to the edge magnetic structure provides another interesting possibility.
The magnitude of the chiral current could be controlled in several ways:
for example, by making an interface to a magnetic material as discussed above, and by applying a magnetic field to the surface of MCIs.  

\section{Summary}\label{sec:summary}
We have numerically investigated the energy spectra and chiral edge current in magnetic Chern insulators with open surfaces or interfaces to magnetic capping layers.
For the magnetic Chern insulator realized in the quarter-filled Kondo lattice model on the triangular lattice, 
we clarified that the edge states for the optimized spin configuration are well described by the junction model with the ferromagnetic interface.
Furthermore, we revealed the close correlation between the total amount of chiral edge current and the energy of the crossing point of the edge states.
We also found that the ferromagnetic arrangement of edge magnetic moments maximizes the total amount of the edge current.
The results suggest that the ferromagnetic edge reconstruction is driven by the optimization of kinetic energy in the edge region,
i.e., a variant of the double-exchange mechanism.
We also discussed the possibility of controlling the chiral edge current via an external magnetic force, by utilizing the sensitivity of the chiral edge current to the local magnetic structure in the edge region.

\ack
R.O. is supported by the Program for Leading Graduate Schools, MEXT, Japan, via the Advanced Leading Graduate Course for Photon Science.
Y.A. acknowledges support from OIST.
This research was supported by Grants-in-Aid for Scientific Research (Grants No. 24340076, 24740221, and 26400339),
the Strategic Programs for Innovative Research (SPIRE), MEXT, and the Computational Materials Science Initiative (CMSI), Japan.

\section*{References}
\providecommand{\newblock}{}

\end{document}